\documentclass[12pt]{article}
\usepackage[english]{babel}
\usepackage[utf8]{inputenc}
\usepackage[T1]{fontenc}
\usepackage{amsmath}
\usepackage{amsfonts}
\usepackage{url}
\usepackage[dvips]{graphicx}
\usepackage{calc}
\usepackage{subfigure}
\usepackage{multirow}

\def\mL{\mathcal{L}}
\def\mH{\mathcal{H}}

\def\bLambda{\bar{\Lambda}}

\def\bA{\mathbf{A}}
\def\bB{\mathbf{B}}
\def\mG{\mathcal{G}}

\begin{document}
	\begin{center}
		{\Large{ \bf Unimodular  Gravity in Covariant
				Formalism}}
		
		\vspace{1em}  J. Kluso\v{n}\textsuperscript{\textdagger} and B. Matou\v{s}\textsuperscript{\textdagger}\textsuperscript{\textdaggerdbl}
		\footnote{Email addresses:
			J. Kluso\v{n}:\ klu@physics.muni.cz, B. Matou\v{s}:\  bmatous@mail.muni.cz}\\
		\vspace{1em} \textsuperscript{\textdagger} \textit{Department of Theoretical Physics and
			Astrophysics, Faculty
			of Science,\\
			Masaryk University, Kotl\'a\v{r}sk\'a 2, 611 37, Brno, Czech Republic}\\
		\vspace{1em} \textsuperscript{\textdaggerdbl} \textit{North-Bohemian Observatory and Planetarium in Teplice, \\
			Kopern\'{i}kova 3062, 415 01, Teplice,
			Czech Republic}\\

	\abstract{In this short note we study unimodular gravity in Weyl-De Donder formalism. We find corresponding Hamiltonian and study consequence of the unimodular constraint on the conjugate covariant momenta. We also find covariant Hamiltonian for Henneaux-Teitelboim unimodular action and study corresponding equations of motion.}

	\end{center}

\section{Introduction and Summary}\label{first}
Unimodular gravity was firstly introduced by A. Einstein in his paper 
\cite{Einstein:1916vd} published in 1916. In this work the unimodular constraint
$\sqrt{-g}=1$ was used as gauge fixing condition 
of general diffeomorphism in order to simplify calculations. Then it was shown in 
\cite{Buchmuller:1988wx,Buchmuller:1988yn} that imposing this condition before the variation of Einstein-Hilbert action leads to the traceless equations of motion. As we review below these equations of motion are classically equivalent to the general relativity equations of motion 
with crucial difference that the cosmological constant appears as integration constant rather than true cosmological constant. This fact brings new hope how to solve cosmological constant problem which was however questioned  in 
\cite{Padilla:2014yea},
\footnote{For review of unimodular gravity, see for example \cite{Jirousek:2023gzr,Carballo-Rubio:2022ofy}.} where it was argued that quantum corrections 
make the cosmological constant ultraviolet sensitive in unimodular gravity as well. On the other hand it is important to stress that no definitive conclusions have been reached yet  regarding this problem and unimodular gravity is still very intensively studied, for some works devoted to unimodular gravity,  see for example
\cite{Alvarez:2023utn,Garay:2023nco,Kehagias:2022mik,Tiwari:2022ctc,Bonder:2022kdw,
	Almeida:2022qld,Kugo:2022dui,Kugo:2022iob,Alonso-Serrano:2021uok,Yamashita:2020ixd,Barvinsky:2019agh,Bufalo:2017tms,Bufalo:2015wda,Gao:2014nia,Jain:2012gc,Smolin:2009ti,Shaposhnikov:2008xb,Finkelstein:2000pg}.

One of the most interesting aspects of unimodular gravity is the number of physical degrees of freedom. Naively, unimodular constraint $\sqrt{-g}=1$ reduces the number of independent components of metric to nine which could suggest that the number of physical degrees of freedom is less than in general relativity. On the other hand unimodular gravity is invariant under restricted diffeomorphism. Taking these two aspects together we find that the number of local physical degrees of freedom is the same as in ordinary general relativity. This fact was proved with the help of the Hamiltonian analysis of unimodular gravity performed in 
\cite{Henneaux:1989zc,Yamashita:2020ixd,Barvinsky:2019agh,Bufalo:2017tms,Bufalo:2015wda,Kluson:2014esa}. On the other hand as was shown in these papers standard analysis of unimodular gravity based on $D+1$ splitting of target space-time is rather non-trivial and shown complexity of the canonical analysis of systems with constraints. 

Then one could ask the question how unimodular gravity could be described in covariant canonical formalism
that is known as Weyl-De Donder theory
\cite{DeDonder,Weyl}. The key point of this formulation is that 
we treat all partial derivatives as equivalent  when we define conjugate momenta. For example, if we have scalar field $\phi$ with Lagrangian density in $D+1$ dimensional space-time equal to $\mL=-\frac{1}{2}\eta^{ab}\partial_a \phi\partial_b\phi-V(\phi)$, we define the conjugate momentum as
\footnote{We define $\eta_{ab}=\mathrm{diag}(-1,1,\dots,1), a,b=0,1,\dots,D$.}
\[\pi^a=\frac{\partial \mL}{\partial \partial_a\phi}=-\eta^{ab}\partial_b\phi \ . \]
Then covariant canonical Hamiltonian
density is defined as
\[\mH=\pi^a\partial_a\phi-\mL=-\frac{1}{2}
\pi_a \eta^{ab}\pi_b+V(\phi) \ .  \]
 Clearly such a form of Hamiltonian density  preserves diffeomorphism invariance of the theory. This approach is known as
multisymplectic field theory, see for example
\cite{Struckmeier:2008zz,Kanatchikov:1997wp,Forger:2002ak}, for review, see \cite{Kastrup:1982qq} and for recent interesting application of this formalism in string theory, see \cite{Lindstrom:2020szt,Kluson:2020pmi}.

It is clear that such covariant canonical formalism is especially suitable for  manifestly covariant theories as for example general relativity. In fact, covariant canonical formalism of general relativity was found long time ago by P. Ho\v{r}ava \cite{Horava:1990ba}. This analysis was recently generalized to the case of $F(R)$ gravity in \cite{Kluson:2020tzn} and further elaborated in \cite{Kluson:2022qxl}.

In this paper we apply this formalism for unimodular theory of gravity in $D+1$ dimensions. This is non-trivial task due to the well known complexity of canonical analysis of unimodular gravity in non-covariant  formalism. Further, it is also very interesting to study this system since it contains primary unimodular constraint and it is non-trivial task how to deal with such systems in covariant canonical formalism. In more details, we include this primary constraint to the action with corresponding Lagrange multiplier. Then we derive corresponding equations of motion. Using these equations of motion we find that the unimodular constraint implies another constraint on the canonical conjugate momenta. Then we show that this constraint is equivalent to the vanishing of the trace of the Christoffel symbols which is characteristic property of unimodular theory of gravity 
\cite{Tiwari:2022ctc}. This is nice and non-trivial result. On the other hand the Lagrange multiplier corresponding to the primary constraint cannot be
determined as in non-covariant canonical formalism by imposing condition of the preservation of the secondary constraint due to the fact that the equations of motion for conjugate momenta are in the form of the divergence of these momenta. For that reason we determine this constraint in the same way as in the Lagrangian formalism when we calculate the trace of the equations of motion. As a result we obtain equations of motion that are traceless and that do not depend on the cosmological constant which is in agreement with the Lagrangian formulation of unimodular gravity. 

 As the second step in our analysis we find covariant canonical formulation 
 of Henneaux-Teitelboim formulation of unimodular gravity
\cite{Henneaux:1989zc}.  In this case we again identify covariant Hamiltonian together with set of primary constraints. Then we consider canonical form of the action and determine corresponding equations of motion. Solving these equations of motion we find that Lagrange multiplier is integration constant. In this case we reproduce results well known from Lagrangian analysis. However we mean that this is nice and interesting application of the covariant canonical analysis to the constraint systems. 

Let us outline our results and suggest possible extension of this work. 
We found covariant Hamiltonian formalism for unimodular gravity. First of all we determined covariant Hamiltonian for general relativity action 
in $D+1$ dimensions where we again introduced variable $f^{ab}=\sqrt{-g}g^{ab}$. At this place we would like to stress an importance of this result since it was not apriori known whether 
$f^{ab}$ is suitable for formulation of gravity in space-time of dimension different from $4$. Then we imposed unimodular constraint using Lagrange multiplier method and then we studied corresponding equations of motion. We found that the consistency of the theory demands that the trace of conjugate momenta is zero. Then we showed that this is characteristic property of unimodular gravity when we pass to Lagrangian formalism. Final we found covariant Hamiltonian for Henneaux-Teltelboim formulation of unimodular gravity. We identified primary constraints of the theory and then we studied equations of motion that follow  from canonical form of the action. We showed that they precisely reproduce Lagrangian equations of motion that is nice consistency check of the covariant canonical formalism. We mean that the analysis presented in this paper suggests that covariant Hamiltonian formalism is very close to Lagrangian formalism and in some situations the covariant
Hamiltonian formalism is more suitable than Lagrangian one, as for example study of thermodynamics properties of horizon \cite{Parattu:2013gwa}. 

It is also clear that there are more systems that could be analysed with the help of covariant canonical formalism. One possibility is to study Weyl invariant gravity in this formalism. Another possibility would be to perform analysis of theories of gravity with higher derivatives where the classical canonical analysis is very complicated, see for example \cite{Kluson:2013hza}. We hope to return to these problems in future. 

This paper is organized as follows. In the next section (\ref{second}) we review properties of unimodular gravity.Then in section (\ref{third}) we proceed to the covariant canonical formulation of this theory. Finally in section (\ref{fourth}) we perform covariant canonical formulation 
of Henneaux-Teltelboim unimodular gravity.

\section{Brief Review of Unimodular Gravity}\label{second}
In this section we review basic facts about unimodular gravity. For recent very nice and more detailed review, see for example 
\cite{Jirousek:2023gzr,Carballo-Rubio:2022ofy}. 
Unimodular gravity is theory with the constraint $\sqrt{-g}=1$.  Clearly such a condition 
has a consequence on allowed differomorphism transformation. In fact, let us consider 
general transformation of coordinates 
\begin{equation}
x'^a=x^a+\xi^a(x) 
\end{equation}
that implies inverse relation 
\begin{equation}
	 x^a=x'^a-\xi^a(x)
\approx x'^a-\xi^a(x')+\mathcal{O}(\xi^2) \ , 
\end{equation}
where $a,b,c=0,1,\dots,D$. Under these transformation the metric $g_{ab}$ transform as
\begin{eqnarray}
g'_{ab}(x)=
g_{ab}(x)
-\partial_c g_{ab}(x)\xi^c(x)
-g_{ac}(x)\partial_b \xi^c(x)-\partial_a \xi^c(x) g_{cb}(x)
\end{eqnarray}
that implies following variation of metric 
\begin{equation}
\delta g_{ab}(x)=g'_{ab}(x)-g_{ab}(x)=
-g_{ac}\partial_b x^c-\partial_a \xi^c
g_{cb}-\partial_c
g_{ab}\xi^c
\nonumber \\
\end{equation} 
so that the variation of the square root of the determinant of metric is equal to
\begin{eqnarray}
	\delta \sqrt{-\det g}=
-(2\partial_a \xi^a-\partial_c g_{ab}g^{ba}\xi^c)\sqrt{-\det g} \ . 
\end{eqnarray}
In case of unimodular gravity this variation should vanish and hence we obtain following condition on  $\xi^a$ in the form 
\begin{eqnarray}\label{restdiff}
\nabla_a \xi^a=\partial_a \xi^a+\frac{1}{2}g^{ac}\partial_d g_{ca}\xi^d=0 \ . 
\nonumber \\
\end{eqnarray}
The most straightforward way how to find an action for unimodular gravity is to 
consider standard Einstein-Hilbert action with an unimodular constraint added
\begin{equation}\label{uniaction}
S=\frac{1}{16\pi}\int d^{D+1}x [\sqrt{-g}(R-2\bar{\Lambda})+\Lambda (\sqrt{-g}-1)]+S_{matt}\ , 
\end{equation}
where $\Lambda$ is Lagrange multiplier whose variation ensures unimodular condition and where 
$\bar{\Lambda}$ is constant. 

 Performing variation of the action (\ref{uniaction}) with respect to 
$g^{ab}$ we obtain following equations of motion 
\begin{equation}\label{eqmotuni}
\frac{1}{16\pi}
(R_{ab}-\frac{1}{2}g_{ab}(R-2\bLambda+\Lambda))=T_{ab} \ , 
\end{equation} 
where $T_{ab}$ is matter stress energy tensor defined as
\begin{equation}
T_{ab}=-\frac{1}{\sqrt{-g}}\frac{\delta S_{matt}}{\delta g^{ab}} \ . 
\end{equation}
The crucial point is that $\Lambda$ is Lagrange multiplier that should be determined as a consequence of the equations of motion. To do this we perform the trace of the equation 
(\ref{eqmotuni}) to express $\Lambda$ as 
\begin{equation}\label{Lambdadet}
	\Lambda=\frac{(1-D)}{1+D}R-\frac{32\pi}{D+1}T+2\bar{\Lambda} \ , \quad 
	T\equiv g^{ab}T_{ab} \ .
\end{equation}
Inserting this result into (\ref{eqmotuni}) we obtain 
\begin{eqnarray}\label{eqmotuni2}
R_{ab}-\frac{1}{D+1}g_{ab}R=16\pi(T_{ab}-\frac{1}{D+1}g_{ab}T) \ . 
\end{eqnarray}
These equations of motion  are trace-free and also most importantly they do not contain any information about  cosmological constant $\bar{\Lambda}$. 

It is important to stress that even equations of motion of general relativity without unimodular constraint imposed split into $9$ trace-free equations of motion and one additional one. To see this consider general relativity equations of motion 
\begin{equation}\label{eqgra}
R_{ab}-\frac{1}{2}g_{ab}(R-2\bLambda)=16\pi T_{ab} \ .
\end{equation}
Taking the trace of this equation we can express $R$ as 
\begin{equation}\label{scaleq}
R=\frac{2}{1-D}(16\pi T-(D+1)\bLambda) \ . 
\end{equation}
Note that with the help of this equation we can rewrite (\ref{eqgra}) into trace-free
form
\begin{equation}
R_{ab}-\frac{1}{D+1}Rg_{ab}=16\pi (T_{ab}-\frac{1}{D+1}Tg_{ab}) \ . 
\end{equation}
However we should again stress that (\ref{scaleq}) determines $R$ as function of trace of matter stress energy tensor and true cosmological constant term in Einstein-Hilbert action while in case of unimodular gravity we express $\Lambda$-which is Lagrange multiplier and not constant, as function of $R,T$ and $\bLambda$, as follows from equation 
(\ref{Lambdadet}).

In order to check equivalence between unimodular gravity and ordinary general relativity we should be able to reproduce equation (\ref{scaleq}) in case of unimodular gravity as well. We can do this by following procedure.  Consider equations of motion (\ref{eqmotuni2}) and rewrite  them into the form
\begin{equation}
	R_{ab}-\frac{1}{2}g_{ab}R=16\pi (T_{ab}-\frac{1}{D+1}g_{ab}T)+\frac{1-D}{2(D+1)}Rg_{ab} \ . 
\end{equation}
Now we apply covariant derivative on both sides of the equations above and using the fact that the covariant derivative of  Einstein tensor $G_{ab}=R_{ab}-\frac{1}{2}g_{ab}R$ is zero we get
\begin{equation}
\frac{1}{D+1}\nabla_b(16\pi T-\frac{1-D}{2}R)=16\pi \nabla^a T_{ab} \ . 
\end{equation}
If we consider ordinary form of matter we obtain that divergence of stress energy tensor is zero as a consequence of \emph{matter equations of motion}. Then the right side of the equation above is zero and the left side can be easily integrated with the result 
\begin{equation}\label{10eq}
R=\frac{2}{1-D}(16\pi T+\Omega) \ , 
\end{equation}
where $\Omega$ now appears as true integration constant rather than the cosmological constant that was imposed in the theory by hand. In other words (\ref{10eq}) is the last equation of motion of unimodular gravity and we fully recovered equivalence with general relativity however keeping in mind that we should still have to impose the condition $\sqrt{-g}=1$ in the course of calculations.

Having performed basic review of unimodular gravity we proceed in the next section to its formulation in the covariant Hamiltonian formalism. 
\section{Covariant Hamiltonian Formalism For $D+1$ dimensional Unimodular Gravity}\label{third}
In this section we find covariant Hamiltonian formalism for unimodular gravity in $D+1$
formalism. 

As usual in the covariant formalism we split the Einstein-Hilbert action into bulk and boundary terms. Since this procedure is well known, see for example 
\cite{Horava:1990ba,Parattu:2013gwa} and also recent generalization to the case of $F(R)$ gravity \cite{Kluson:2020tzn} we write immediately final result
\begin{eqnarray}\label{Lden}
&&\mL=\mL_{bulk}+\mL_{surf} \ , \nonumber \\
&&\mL_{bulk}=\frac{1}{16\pi}\sqrt{-g}[\Gamma^h_{dk}\Gamma^k_{gh}g^{gd}-
\Gamma^f_{fk}\Gamma^k_{gh}g^{gh}]+\nonumber \\
&&+\frac{1}{16\pi}\bLambda \sqrt{-g}+
\frac{1}{16\pi}\lambda (\sqrt{-g}-1)\equiv 
 \nonumber \\
&&\equiv\mL_{quad}+\frac{1}{16\pi}\bLambda \sqrt{-g}+
\frac{1}{16\pi}\lambda (\sqrt{-g}-1) \ , \nonumber \\
&&\mL_{surf}=\frac{1}{16\pi}\partial_j[\sqrt{-g}(g^{ik}\Gamma_{ik}^j-g^{ij}\Gamma_{ik}^k)]\ ,  \nonumber \\
\end{eqnarray}
where $\Gamma^a_{bc}$ are Christoffel symbols 
\begin{equation}
\Gamma^a_{bc}=\frac{1}{2}g^{ad}(\partial_b g_{dc}+\partial_c g_{db}-\partial_c g_{ab}) \ ,
\end{equation}
and where $\bLambda$ is cosmological constant. 
Note that the presence of the term with Lagrange multiplier allows us to treat all components of metric as independent. 

Now we are ready to proceed to the covariant Hamiltonian formulation of this theory.
The main idea of this formalism is to treat all derivatives of dynamical variables on the equal footing \cite{DeDonder,Struckmeier:2008zz,Horava:1990ba} which is sharp contrast with the standard canonical formalism where the time coordinate has exceptional meaning. This is very attractive idea especially in the context of generally covariant theories since sometimes it is very difficult to perform $D+1$ splitting of targe-space time and corresponding dynamical fields. In case of covariant canonical formalism of gravity we define conjugate momenta $M^{cmn}$ to $g_{mn}$ in the following way
\begin{equation}
M^{cmn}=\frac{\partial \mL_{bulk}}{\partial \partial_c g_{mn}} \ . 
\end{equation}
Note that the momenta are defined by bulk part of the Lagrangian density only as follows from 
the fact that equations of motion are derived by variation of the action when we fix metric and its derivative on the boundary, for careful discussion see \cite{Parattu:2013gwa}.

Then from (\ref{Lden}) we  obtain
\begin{eqnarray}
&&M^{cmn}
=\frac{1}{32\pi}\sqrt{-g}
[g^{mk}\Gamma^c_{kd}g^{dn}+g^{nk}\Gamma^c_{kd}g^{dm}-\nonumber \\
&&-g^{mn}\Gamma_{gh}^cg^{gh}-\Gamma^f_{fk}
(g^{km}g^{cn}+g^{kn}g^{cm})+
g^{mn}g^{ck}\Gamma^f_{fk}]
\nonumber \\
\end{eqnarray}
using
\begin{eqnarray}
&&\frac{\delta \Gamma^k_{gh}}{\delta \partial_cg_{mn}}=
\frac{1}{4}(g^{ks}\delta_g^c(\delta_s^m\delta_h^n+\delta_s^n\delta_h^m)+
\nonumber \\
&&+
g^{ks}\delta_h^c(\delta_s^m\delta_g^n+\delta_s^n\delta_g^m)-
g^{ks}\delta_s^c(\delta_g^m\delta_h^n+\delta_g^n\delta_h^m)) \ \ \nonumber \\
\end{eqnarray}
Then we could formulate covariant Hamiltonian formalism using canonical variales
$g_{ab}$ and $M^{cab}$. However it turns out that the situation is much simpler when we
introduce an alternative set of variables 
\cite{Horava:1990ba,Parattu:2013gwa} that are defined as 
\begin{equation}\label{deff}
f^{ab}=\sqrt{-g}g^{ab} \ . 
\end{equation}
Then it is easy to see that the conjugate momenta are defined by chain rule 
\begin{eqnarray}
N^c_{ \ ab}=\frac{\partial \mL_{quad}}{\partial \partial_c f^{ab}}
=\frac{\partial L_{quad}}{\partial(\partial_d g_{mn})}\frac{\partial (
	\partial_d g_{mn})}{\partial(\partial_c f_{ab})} \ . 
\nonumber \\
\end{eqnarray}
From (\ref{deff}) we see that   $f^{ab}$ and $g_{mn}$ are related by point transformations so that 
\begin{equation}
\partial_d g_{mn}=\frac{\partial g_{mn}}{\partial f^{ab}}\partial_d f^{ab} \ . 
\end{equation}
 Then  we have
\begin{eqnarray}
\frac{\partial (\partial_d g_{mn})}{\partial (\partial_c f^{ab})}=
\frac{\partial g_{mn}}{\partial f^{ab}}\delta_d^c \nonumber \\
\end{eqnarray}
and finally 
\begin{eqnarray}
N^c_{ \ ab}=\frac{\partial \mL_{quad}}{\partial(\partial_c g_{mn})}(-g_{mk}B^{kl}_{ \ ab}g_{ln})
 \ , \nonumber \\
\end{eqnarray}
where 
\begin{eqnarray}
B^{kl}_{ \ ab}=\frac{\delta g^{kl}}{\delta f^{ab}}
=(-f)^{-\frac{1}{D-1}}\left(\frac{1}{2}(\delta^k_a\delta^l_b+\delta^l_a\delta^k_b)-\frac{1}{D-1}
f^{kl}f_{ab}\right) \  , \nonumber \\
\end{eqnarray}
where we used the fact that
\begin{equation}
-\det f\equiv -f=(-g)^{\frac{D+1}{2}}(-g)^{-1} 
\end{equation}
and consequently 
\begin{equation}
\sqrt{-g}=(-f)^{\frac{1}{D-1}}  \ , \quad  
g^{ab}=(-f)^{-\frac{1}{D-1}}f^{ab}  \ . 
\end{equation}
Then using previous form of $M^{cmn}$ we obtain
\begin{eqnarray}\label{defN}
&&N^c_{ \ ab}=\frac{\partial \mL_{quad}}{\partial(\partial_c g_{mn})}(-g_{mk}B^{kl}_{ \ ab}g_{ln})=\nonumber \\
&&=-\frac{1}{32\pi}[2\Gamma_{ab}^c-\Gamma^f_{fa}\delta^c_b-
\Gamma^f_{fb}\delta^c_a] \ . \nonumber \\
\end{eqnarray}
Note that this relation does not depend on the number of space-time dimensions. Then in order to find corresponding Hamiltonian we should find inverse relation between $\Gamma^a_{bc}$ and $N^a_{bc}$. Let us presume  
that it has the form
\begin{equation}\label{relpre}
\Gamma_{ab}^c=\bA N^c_{ab}+\bB(N_{da}^d\delta^c_b+N_{bd}^d\delta_a^c) \ . 
\end{equation}
Inserting (\ref{defN}) into (\ref{relpre}) we obtain
\begin{eqnarray}
&&N^c_{ab}=-\frac{1}{32\pi} 
(2\bA N_{ab}^c+2\bB(N^d_{da}\delta^c_b+
N_{bd}^d\delta_a^c)-\nonumber \\
&&-(\bA +\bB(D+2))N^f_{fa}\delta_b^c-(\bA +\bB(D+2))N^f_{fb}\delta_a^c)
\nonumber \\
\end{eqnarray}
using $\Gamma^f_{fa}=(\bA +\bB(D+2))N^f_{fa}$. Comparing left and right side we obtain that $\bA$ and $\bB$ are equal to
\begin{equation}
\bA=-16\pi \ , \quad 
\bB=-\frac{\bA}{D} \ .
\end{equation}
 Then it is easy to find kinetic term of covariant Hamiltonian for $D+1$ dimensional unimodular gravity in the form 
\begin{eqnarray}\label{defkin}
\mH_{kin}=\partial_c f^{ab}N_{ab}^c-\mL_{quad}=
16\pi
\left[N^b_{cd}f^{da}N_{ab}^c-\frac{1}{D}N^r_{ra}f^{ab}N^s_{sb}\right] \ , 
\nonumber \\
\end{eqnarray}
where we used the fact that
\begin{eqnarray}
\partial_c f^{ab}=\partial_c \sqrt{-g}g^{ab}+
\sqrt{-g}\partial_c g^{ab}=
\Gamma_{dc}^d f^{ab}-\Gamma^a_{cd}f^{db}-
\Gamma^b_{dc}f^{da}
\nonumber \\
\end{eqnarray}
together with the condition $\nabla_c g^{ab}=0$ that implies 
\begin{eqnarray}
\partial_c \sqrt{-g}=\Gamma^d_{dc}\sqrt{-g} \ , \quad 
\partial_c g^{ab}=-(\Gamma^a_{cd}g^{db}+
\Gamma^b_{cd}g^{da}) \ .
\nonumber \\
\end{eqnarray}
The final form of the covariant Hamiltonian for unimodular gravity 
contains terms with the unimodular constraint and true cosmological constant $\bLambda$. Then the phase-space form of the action has the form
\begin{equation}
S=\int d^{D+1}x (N^c_{ab}\partial_c f_{ab}-\mH_{kin}-\frac{1}{16\pi}(-f)^{\frac{1}{D-1}}\bLambda-\frac{1}{16\pi}\lambda ((-f)^{\frac{1}{D-1}}-1)) \ , 
\end{equation}
where $\lambda$ is Lagrange multiplier corresponding to unimodular constraint. From the action above we  determine corresponding equations of motion by performing variation with respect to $f^{ab},N^c_{ab}$ and $\lambda$
\begin{eqnarray}
&&\delta S=\int 
d^{D+1}x
(\delta N^c_{ab}\partial_c f_{ab}+N^c_{ab}\partial_c \delta f_{ab}-\nonumber \\
&&-\frac{\delta \mH_{kin}}{\delta N^c_{ab}}\delta N^c_{ab}-
\frac{\delta \mH_{kin}}{\delta f^{ab}}\delta f^{ab}-\nonumber \\
&&-\frac{1}{16\pi(D-1)}(\lambda+\bLambda)( -f)^{\frac{1}{D-1}}\delta f^{ab}f_{ab}-\delta \lambda ((-f)^{\frac{1}{D-1}}-1))=0
\nonumber \\
\end{eqnarray}
that implies following equations of motion
\begin{eqnarray}
&&\partial_c f^{ab}=\frac{\delta \mH}{\delta N^c_{ab}} \ ,  \quad (-f)^{\frac{1}{D-1}}-1=0 \ , \nonumber \\
&&-\partial_c N^c_{ab}=\frac{\delta \mH}{\delta f^{ab}}+\frac{\lambda}{16\pi(D-1)}
(-f)^{\frac{1}{D-1}}f_{ab}+\frac{\bLambda}{16\pi(D-1)}(-f)^{\frac{1}{D-1}}f_{ab} \ , \nonumber \\
\end{eqnarray}
or explicitly 
\begin{eqnarray}\label{eqfab}
&&\partial_c f^{ab}=16\pi
[N^a_{cd}f^{db}+N^b_{cd}f^{da}-\frac{1}{D}
(f^{bd}N_{sd}^s \delta_c^a+f^{ad}N_{sd}^s\delta_c^b)] \ , 
\nonumber \\
&&-\partial_c N^c_{ab}=\frac{16\pi}{2}(
N^d_{ca}N^c_{bd}+N^d_{cb}N^c_{ad})-\nonumber \\
&&-\frac{16\pi}{ D}N^r_{ra}N^s_{sb}
+\frac{\lambda}{16\pi(D-1)}
(-f)^{\frac{1}{D-1}}f_{ab}+\frac{\bLambda}{16\pi(D-1)}(-f)^{\frac{1}{D-1}}f_{ab} \ , 
\nonumber \\
&&(-f)^{\frac{1}{D-1}}-1=0 \ . \nonumber \\
\end{eqnarray}
Taking the trace of the second equation we can determine $\lambda$ as
\begin{eqnarray}
\lambda=\frac{16\pi(D-1)}{(D+1)}(-\partial_c N^c_{ab}f^{ab}-16\pi N^d_{ca}f^{ab}N^c_{bd}+\frac{16\pi}{ D}
N^r_{ra}f^{ab}N^s_{sb})-\bLambda \ ,
\nonumber \\
\end{eqnarray}
where we have took into account the equation on the fourth line in (\ref{eqfab}). 
Then the  equations of motion for $N^c_{ab}$ 
have the form
\begin{eqnarray}
&&-\partial_c N^c_{ab}=\frac{16\pi}{2}(
N^d_{ca}N^c_{bd}+N^d_{cb}N^c_{ad})-\frac{16\pi}{ D}N^r_{ra}N^s_{sb}+\nonumber \\
&&+
\frac{1}{(D+1)}(-\partial_j N^j_{ik}f^{ik}-16\pi N^d_{ci}f^{ik}N^c_{kd}+\frac{16\pi}{ D}
N^r_{ri}f^{ik}N^s_{sk})
f_{ab} \ .  \nonumber \\
\end{eqnarray}
Clearly this equation is traceless and all dependence on the cosmological constant $\bLambda$ disappears which is an essence of unimodular gravity. 

On the other hand one let us try to calculate the trace of the first equation that gives
\begin{equation}
	\partial_c f^{ab}f_{ab}=16\pi
	[N^a_{cd}f^{db}+N^b_{cd}f^{da}-\frac{1}{D}
	(f^{bd}N_{sd}^s \delta_c^a+f^{ad}N_{sd}^s\delta_c^b)]f_{ba} 
\end{equation}	
	that can be simplified into the form 
\begin{equation}
\partial_c f=32\pi[\frac{D-1}{D}]N^s_{sc} \ .  \nonumber \\
\end{equation}
Now taking into account unimodular constraint we immediately get the condition
\begin{equation}\label{Nssc}
	N^s_{sc}=0 \  
\end{equation}
that can be interpreted as secondary constraint. On the other hand the condition (\ref{Nssc}) seems to be too strong so that we should discuss it in more details. 

We begin with the recapitulation that unimodular gravity in the covariant Hamiltonian formalism is described by canonical conjugate variables $f^{ab},N^c_{ab}$ that are restricted by unimodular condition together with (\ref{Nssc}). In order to find proper interpretation of the constraint (\ref{Nssc}) it is instructive to derive general relativity variables from  $f^{ab},N_{ab}^c$. As the first step let us consider linear combination of $N_{ab}^c$ that we denote as $\Gamma_{ab}^c$ and which is given by following prescription 
\begin{equation}\label{GammaN}
\Gamma_{ab}^c=-16\pi N^c_{ab}+\frac{16\pi}{D}(N_{da}^d\delta^c_b+N_{bd}^d\delta_a^c) \ .
\end{equation}
This can be always done and we should again stress that $\Gamma_{ab}^c$ is not related to $f^{ab}$ at all. Clearly $\Gamma_{ab}^c=\Gamma^c_{ba}$. Then we define covariant derivative  where $\Gamma_{ab}^c$ are coefficients of connection. Let us further define  $g^{ab}$ and its inverse $g_{ab}$ in the following way
\begin{eqnarray}
g^{ab}
=f^{ab}(-f)^{\frac{1}{1-D}} \ , \quad 
g_{ab}
=f_{ab}(-f)^{\frac{1}{D-1}} \ .
\end{eqnarray}
Let us then define covariant derivative of $g^{ab}$ as
\begin{equation}
	\nabla_c g^{ab}=\partial_c g^{ab}+\Gamma^a_{cd}g^{db}+
	\Gamma^b_{cd}g^{da}  \ , 
\end{equation}
that, using (\ref{GammaN}),  takes the form
\begin{eqnarray}
&&\nabla_c g^{ab}
=(-f)^{\frac{1}{1-D}}\times \nonumber \\
&&\times 
[\partial_c f^{ab}-16\pi N^a_{cd}f^{db}-16\pi N^b_{cd}f^{da}
+\frac{16\pi}{D}f^{bd}N^r_{dr}\delta^a_c+\frac{16\pi}{D}N^r_{dr}f^{da}\delta^b_c]=0 \ , 
\nonumber \\
\end{eqnarray}
where we used  the first equation in (\ref{eqfab})
that also implies $\partial_c f^{mn}f_{mn}=32\pi \frac{D-1}{D}N^s_{sc}$.
Now thanks to the equation $\nabla_c g^{ab}=0$ we can express $\Gamma^a_{bc}$ in the form of Christoffel symbols
\begin{equation}\label{defGamma}
	\Gamma^a_{bc}=\frac{1}{2}g^{ad}(\partial_b g_{dc}+\partial_c g_{db}-\partial_d g_{bc}) \ . 
\end{equation}

On the other hand 
let us return to the relation between $\Gamma^a_{bc}$ and $N^a_{bc}$ that takes the form 
\begin{equation}
	\Gamma^f_{fa}=-\frac{32\pi}{D}N^f_{fa} \ 
\end{equation}
so that condition that $N^s_{sa}=0$ implies 
\begin{equation}
	\Gamma^s_{sa}=0 \ . 
\end{equation}
On the other hand from (\ref{defGamma}) we obtain 
\begin{equation}
	\Gamma^f_{fc}=\frac{1}{2}g^{fd}\partial_c g_{df}=
	\partial_c \det g=0
\end{equation}
so that condition $N^s_{sc}=0$ is equivalent to unimodular condition. It is important to stress that the fact that unimodular constraint implies $\Gamma^s_{sa}=0$ has not been appreciated too much with exception of recent interesting paper \cite{Tiwari:2022ctc}
where it was stressed that the equivalence between general relativity and unimodular gravity 
is non-trivial. Rather, it was argued there 
that the natural geometry for unimodular relativity is equiprojective
geometry \cite{Thomas}. We also see that the condition $N^s_{sa}=0$ emerges  naturally in the covariant canonical formalism of unimodular gravity.



\section{Covariant Form of Unimodular Gravity}\label{fourth}
In this section we perform covariant canonical formalism for Henneaux-Teitelboim  formulation of unimodular gravity that has the form 
\begin{equation}
S=\frac{1}{16\pi}\int d^{D+1}x\sqrt{-g}[R+\lambda(\sqrt{-g}-\partial_a \tau^a)] \ , 
\end{equation}
where $\tau^a$ is vector density and $\lambda$ is Lagrange multiplier. 
Now the equations of motion for $\lambda$ implies 
\begin{equation}\label{sqrtgtau}
\sqrt{-g}-\partial_a \tau^a=0
\end{equation}
while equation of motion for $\tau^a$ leads to 
\begin{equation}
\partial_a \lambda=0 \ . 
\end{equation}
It is clear that the covariant Hamiltonian formulation of this theory is almost the same 
as in previous case with difference that there is momentum conjugate to $\tau^a$. Writting
$\partial_a \tau^a=\partial_b\tau^a\delta_a^b$ we obtain momentum conjugate to $\tau^a$ to be equal to
\begin{equation}
p_a^b=\frac{\delta \mL}{\delta \partial_b \tau^a}=
-\frac{1}{16\pi}\lambda \delta_a^b 
\end{equation}
however this can be interpreted as  primary constraints of the theory
\begin{equation}
\mG^b_a\equiv p^b_a+\frac{1}{16\pi}\lambda \delta^b_a
\ . 
\end{equation}
In fact, the bare Hamiltonian is defined as
\begin{eqnarray}
&&\mH_B=p_a^b\partial_b\tau^a+\partial_c f^{ab}N_{ab}^c-\mL=
\nonumber \\
&&=16\pi
[N^b_{cd}f^{da}N_{ab}^c-\frac{1}{D}N^r_{ra}f^{ab}N^s_{sb}]-\frac{1}{16\pi}\lambda (-f)^{\frac{1}{D-1}}
\nonumber \\
\end{eqnarray}
and we see that the dependence on momenta $p_\mu^\nu$ is missing. For that reason we should consider Hamiltonian with primary constraints included
\begin{eqnarray}
&&\mH_T=16\pi
[N^b_{cd}f^{da}N_{ab}^c-\frac{1}{D}N^r_{ra}f^{ab}N^s_{sb}]
-\nonumber \\
&&\frac{1}{16\pi}\lambda (-f)^{\frac{1}{D-1}}+\Gamma^a_b
(p^b_a+\frac{1}{16\pi}\lambda \delta_a^b)
\nonumber \\
\end{eqnarray}
and consider corresponding equations of motion that arise from the variation of the canonical form of the action
\begin{eqnarray}
&&S=\int d^{D+1}x (\partial_c f^{ab}N^c_{ab}+p^a_b\partial_a \tau^b-
16\pi
[N^b_{cd}f^{da}N_{ab}^c-\frac{1}{D}N^r_{ra}f^{ab}N^s_{sb}]+
\nonumber \\
&&+\frac{1}{16\pi}\lambda (-f)^{\frac{1}{D-1}}+\Gamma^a_b
(p^b_a+\frac{1}{16\pi}\lambda \delta_a^b))
\nonumber \\
\end{eqnarray}
so that the equations of motion have the form
\begin{eqnarray}\label{eqHT}
&&\partial_c f^{ab}=16\pi
[N^a_{cd}f^{db}+N^b_{cd}f^{da}-\frac{1}{D}
(f^{bd}N_{sd}^s \delta_c^a+f^{ad}N_{sd}^s\delta_c^b)] \ , 
\nonumber \\
&&-\partial_c N^c_{ab}=\frac{16\pi}{2}(
N^d_{ca}N^c_{bd}+N^d_{cb}N^c_{ad})-\frac{16\pi}{ D}N^r_{ra}N^s_{sb}+\frac{\lambda}{(D-1)}
(-f)^{\frac{1}{D-1}}f_{ab} \ , \nonumber \\
&& (-f)^{\frac{1}{D-1}}+\Gamma^a_a=0 \ ,  \quad \partial_b \tau^a+ \Gamma^a_b=0 \ , \quad \partial_a p^a_b=0 \ ,  \quad 
p_a^b+\frac{1}{16\pi}\lambda \delta_a^b=0 \ . 
\nonumber \\
\end{eqnarray}
If we combine the first and the second equation on the third line we find 
\begin{equation}
(-f)^{\frac{1}{D-1}}=\partial_a\tau^a 
	\end{equation}
that has exactly the same form as equation (\ref{sqrtgtau}). We further perform 
partial derivative of the fourth equation on the third line and we obtain 
\begin{equation}
\partial_b p^b_a=-\frac{1}{16\pi}\partial_a \lambda 
\end{equation}
that using the third equation on the same line implies that 
\begin{equation}
\partial_a \lambda=0 \ . 
\end{equation}
This equation also shows that $\lambda$ is constant and it can be interpreted as integration constant. Then it can be argued in the same way as in the previous section that the  equations 
(\ref{eqHT}) 
are equivalent to the Lagrangian equations of Henneaux-Teitelboim gravity. In other words, covariant Hamiltonian description of Henneaux-Teiltelboim gravity is equivalent to corresponding Lagrangian description which is nice consistency check.  

{\bf Acknowledgement:}

The work of JK is supported by the grant “Dualitites and higher order derivatives” (GA23-06498S) from the Czech Science Foundation (GACR).

\end{document}